\documentstyle[preprint,aps]{revtex}
\preprint{IMSc-96/06/18}
\begin{document}
\draft 
\title{Atmospheric neutrinos with three flavor mixing} 
\author
{Mohan Narayan, G. Rajasekaran} 
\address
{Institute of Mathematical Sciences, Madras 600 113, India.}
\author
{S. Uma Sankar}
\address
{Department of Physics, I.I.T., Powai, Bombay 400 076, India.}
\date{\today}
\maketitle

\begin{abstract}
We analyze the atmospheric neutrino data in the context 
of three flavor neutrino oscillations taking account of 
the matter effects in the earth.  
With the  hierarchy among the vacuum 
mass eigenvalues $\mu_3^2 \gg \mu_2^2 \geq \mu_1^2$, the solution of
the atmospheric neutrino problem depends on $\delta_{31}=\mu_3^2 -
\mu_1^2$ 
and the $13$ and $23$ mixing angles $\phi$ and $\psi$. 
Whereas the sub-GeV atmospheric neutrino data imposes only
a lower limit on $\delta_{31}  > 10^{-3} \ eV^2$,
the zenith angle dependent suppression observed in the multi-GeV 
data limits $\delta_{31}$ from above also. The allowed regions of
the parameter space are strongly constrained by the multi-GeV data.
Combined with our earlier solution to the solar neutrino problem
which depends on $\delta_{21}= \mu_2^2-\mu_1^2$ and the $12$ and $13$
mixing angles $\omega$ and $\phi$, we have obtained the ranges of values
of the five neutrino parameters which solve both the solar and the
atmospheric neutrino problems simultaneously.
\end{abstract}
\vspace{0.5cm}
\narrowtext

\newpage 
\section{Introduction}

In recent years, deep underground 
detectors have been measuring the flux        
of neutrinos produced in the atmosphere. (Unless explicitly stated,
we call neutrinos and anti-neutrinos collectively as neutrinos).
These neutrinos are produced from the decays
of $\pi^{\pm}$ and $K^{\pm}$ mesons which, in turn, are 
produced by cosmic ray interactions with the atmosphere. One
expects the number of muon neutrinos   to be twice 
the number of electron neutrinos.
The predictions
of detailed Monte Carlo calculations confirm that the ratio of the
flux of muon neutrinos 
$\Phi_{\nu_{\mu}}$ to the flux of electron neutrinos 
$\Phi_{\nu_e}$ is about $2$
\cite{honda,agrawal}. The predicted neutrino fluxes from different
calculations
differ significantly from each other (by as much as $30 \%$),
but the predictions for the ratio of the fluxes 
$\Phi_{\nu_{\mu}}/\Phi_{\nu_e}$  
are in good agreement with each other. The large water Cerenkov 
detectors Kamiokande and IMB \cite{imb} have measured this ratio 
and have found it to be about half of what is predicted \cite{fn1}. 
The experimental results are presented in the form of a double ratio 
\begin{equation}
R = \frac{\left(\frac{N_{\nu_{\mu}}}{N_{\nu_e}}\right)_{obs}}  
{\left(\frac{N_{\nu_{\mu}}}{N_{\nu_e}}\right)_{MC}}
= \frac{r_{obs}}{r_{MC}}. \label{eq:defR}  
\end{equation}

Kamiokande collaboration have presented their results for neutrinos
with energy less than $1.33$ GeV (sub-GeV data) \cite{hirata} 
and for neutrinos with energy greater then $1.33$ GeV (multi-GeV
data) \cite{fukuda}. For the sub-GeV data,     
$R=0.60^{+0.07}_{-0.06}\pm0.05$ and for the multi-GeV data, after  
averaging over the zenith angle,   
 $R=0.57^{+0.08}_{-0.07}\pm0.07$. The value of $R$
has no significant zenith angle dependence for the sub-GeV data.  
However, for the multi-GeV data, $R$ is smaller for 
large values of zenith angle (upward going neutrinos) and is larger
for small values of zenith angle (downward going neutrinos). 
Kamiokande have analyzed their data assuming that
the smaller observed value of 
$R$ is caused by neutrino oscillations.
They have done two independent analyses, one assuming 
two flavor oscillations between $\nu_\mu \leftrightarrow \nu_e$ 
and the other assuming two flavor oscillations between
$\nu_\mu \leftrightarrow \nu_\tau$. For both the cases,
they obtain a mass-squared difference 
$\delta m^2 \sim 10^{-2} eV^2$ and 
a mixing angle nearly $45^0$. Since the upward going neutrinos 
travel large distances inside the earth before entering the  
detector, matter effects may be important for these, especially
at higher energies \cite{sandip,gvdkvl}.
Therefore one must take matter effects into consideration while
analyzing $\nu_\mu \leftrightarrow \nu_e$ oscillations. 

Atmospheric neutrino problem was analyzed in the three flavor 
neutrino oscillation framework previously. In Ref. \cite{jim}
the sub-GeV data were analyzed under the assumption that one of  
the mass differences is much smaller than the other. The matter 
effects due to the passage of neutrinos through the earth were 
included and the allowed values of neutrino parameters were obtained. 
Various other authors have analyzed, in the context of three flavor 
oscillations, accelerator and 
reactor data in conjunction with the sub-GeV 
atmospheric data \cite{sruba} or the multi-GeV data with zenith angle 
dependence included \cite{bilenky}.  However, in both these cases the
earth matter effects were not taken into account.

Several authors have attempted a simultaneous solution of the solar
and atmospheric neutrino problems 
in three flavor oscillation scenarios.
The solution in Ref. \cite{giunti} assumes maximal mixing between all
the three flavors and derives the constraints on the mass differences.
The solution thus obtained restricts the mass difference to be very 
small $(\sim 10^{-10} \ eV^2)$ and is somewhat fine-tuned. Other 
solutions have assumed the mass hierarchy that was considered  
in Ref. \cite{jim} and obtained the allowed regions in the neutrino
parameter space \cite{harley,fogli,anjan,narayan}. However, all these
analyses were based on the sub-GeV atmospheric neutrino data.
The multi-GeV data and its zenith angle dependence were not included.
In a recent analysis of the multi-GeV 
data earth matter effects and the zenith angle dependence were 
taken into account \cite{yasuda}. However, this analysis considered   
the atmospheirc neutrino problem only. 

In this paper, we analyze the Kamiokande multi-GeV data in the 
framework of three flavor neutrino oscillations. The solar neutrino
data and the sub-GeV data of Kamiokande were analyzed earlier in the
same framework \cite{narayan}. We only assume that, of the two mass
differences in the problem, the smaller one is chosen to solve the 
solar neutrino problem and the larger one will be relevant for the 
atmospheric neutrino problem. We do not, apriori, make any assumption
about the form of the mixing matrix. Our starting point is the same as  
that in Ref. \cite{jim}. We 
include the matter effects due to the passage
of neutrinos through the earth. We take the allowed values of neutrino  
parameters from our earlier work as inputs. The zenith angle dependent
suppression of multi-GeV data places strong further constraints on the 
parameter space
allowed by the solar neutrino problem and the sub-GeV atmospheric 
neutrino data. In section II, we define our theoretical 
framework and derive the oscillation probabilities. In section III,
we describe the comparison of the theoretical predictions with the 
Kamiokande
data and determine the allowed regions of the neutrino parameters. 
In section IV we discuss our results and present our conclusions. 
A brief discussion of the recent LSND 
result \cite{lsnd} is also included.

\section{Theory}

Here we describe the three flavor neutrino mixing and 
calculate the probability for a neutrino produced as $\nu_\alpha$
to be detected as $\nu_\beta$, where $\alpha$ and $\beta$ are
flavor indices. The flavor eigenstates are related to the 
mass eigenstates by 
\begin{equation}
\left[ \begin{array}{c} 
\nu_e \\ \nu_\mu \\ \nu_\tau \end{array} \right] 
= U^v 
\left[ \begin{array}{c} 
\nu_1^v \\ \nu_2^v \\ \nu_3^v \end{array} \right] 
\end{equation}
where the superscript `v' denotes vacuum. 
The unitary matrix $U^v$ can be parametrized as
\begin{equation}
U^v = U^{23} (\psi) \times U^{phase} \times U^{13} (\phi)
\times U^{12} (\omega),
\label{eq:defUv}
\end{equation}
where $U^{ij} (\theta_{ij})$ is the two flavor mixing matrix between 
the ith and jth mass eignestates with the mixing angle $\theta_{ij}$.
For simplicity, we neglect the CP violation and set 
$U^{phase} = I$.

In the mass basis
the mass-squared matrix is diagonal and can be taken to be
\begin{equation}
M_0^2 = \left[ \begin{array}{ccc}
\mu_1^2 & 0 & 0 \\ 0 & \mu_2^2 & 0 \\ 0 & 0 & \mu_3^2
\end{array} \right]
= \mu_1^2 I + \left[ \begin{array}{ccc}
0 & 0 & 0 \\ 0 & \delta_{21} & 0 \\ 0 & 0 & \delta_{31}
\end{array} \right],
\end{equation}
where $\delta_{21} = \mu_2^2 - \mu_1^2$ and $\delta_{31} =  
\mu_3^2 - \mu_1^2$. 
Without loss of generality we can take $\mu_3 > \mu_2 > \mu_1$
so that both $\delta_{21}$ and $\delta_{31}$ are positive.
For extreme relativistic neutrinos, the
oscillation probability depends only on the mass-squared differences 
so we ignore the first term $\mu_1^2 I$.

In the flavor basis, the mass-squared matrix is non-diagonal and
is given by
\begin{equation}
M_v^2 = U^v M_0^2 \left( U^v \right)^{\dag}.   
\end{equation}
For the propagation of neutrinos through the earth, we need to take
the matter effects into account. The charged current scattering 
between electrons and $\nu_e$ induces an effective mass-squared term
for $\nu_e$ which is of the form $A = 2 \sqrt{2} G_F N_e E$, where 
$N_e$ is the number density of electrons and E is the neutrino energy
 \cite{wolfst}. 
This term is present only for the $e-e$ element in the flavor basis
so the mass-squared matrix including the matter effects is
\begin{equation}
M_m^2 = M_v^2 + \left[ \begin{array}{ccc}
A & 0 & 0 \\ 0 & 0 & 0 \\ 0 & 0 & 0 \end{array} \right].
\label{eq:Mmsq}
\end{equation}
$M_m^2$ is a hermitian matirx and can be diagonalized by 
unitary matrix $U^m$, which relates the flavor eigenstates to 
matter dependent mass eigenstates
\begin{equation}
\left[ \begin{array}{c} 
\nu_e \\ \nu_\mu \\ \nu_\tau \end{array} \right] 
= U^m 
\left[ \begin{array}{c} 
\nu_1^m \\ \nu_2^m \\ \nu_3^m \end{array} \right] 
\end{equation}
where the superscript `m' stands for matter.  
We denote the matter dependent mass eigenvalues as 
$m_1$, $m_2$ and $m_3$.
The matter dependent mixing matrix $U^m$ can be parametrized 
in terms of three mixing angles in a manner similar to that of 
$U^v$ as given in eq.(\ref{eq:defUv}). The matter dependent mass
eigenvalues and mixing angles can be obtained, in terms of the 
vacuum parameters and $A$, by solving the eigenvalue problem of
$M_m^2$ in eq.(\ref{eq:Mmsq})
 
The distance scales and the energy scales in the solar neutrino 
problem and the atmospheric neutrino problem are very different.
Therefore, one needs two distinct mass scales to solve the solar
and the atmospheric neutrino problems simultaneously. To satisfy the
constraints from all the three solar neutrino experiments, one
must choose $\delta_{21} \sim 10^{-5} \ eV^2$, which is roughly
the matter term $A$ for the solar neutrinos due to their passage
through the sun. As we will shortly see, the sub-GeV atmospheric
neutrino data impose a constraint $\delta_{31} \geq 10^{-3} \ eV^2$.
Hence $\delta_{31} \gg \delta_{21}$. 
If $\delta_{21} \sim 10^{-5} \ eV^2$,
the oscillation length corresponding to it, even for the minimum 
of the atmospheric neutrino energies, is of the order of the diameter 
of the earth. In the expression for the oscillation probability,
$\delta_{21}$ can  be set to zero. Therefore the oscillations are 
dependent on only one mass difference $\delta_{31}$ in the 
atmospheric neutrino problem. 

In the approximation of neglecting $\delta_{21}$, it is 
straightforward to show that the oscillation probability is
independent of the mixing angle $\omega$. Including matter effects
does not alter this conclusion. Note, however, that neglecting
$\delta_{21}$ does not reduce the problem to an effective two
flavor mixing. The three flavor nature of the problem is reflected  
by the fact that the oscillation probability is a function of the
mass difference $\delta_{31}$ and {\it two} mixing angles $\phi$
and $\psi$. In the case of an effective two flavor mixing, the
oscillation probability is dependent on one mixing angle only.
When the matter effects are included the mixing angle $\psi$
remains unaffected but the angle $\phi$ becomes matter dependent 
\cite{jim},
\begin{equation}
\tan 2 \phi_m = \frac{\delta_{31} \sin 2 \phi}{
               \delta_{31} \cos 2 \phi - A}.
\label{eq:t2phim}
\end{equation}
The mass eigenvalue of $\nu_1^m$ remains $0$ (actually it is of the
order of $\delta_{21}$ which we are neglecting here). The other two 
matter dependent mass eigenvalues are given by 
\begin{eqnarray}
m_2^2 & = &  \frac{1}{2} \left[ \left(\delta_{31} + A \right)
           - \sqrt{\left(\delta_{31} \cos 2 \phi - A \right)^2 +
                   \left(\delta_{31} \sin 2 \phi \right)^2} \right] 
\label{eq:m2sq}  \\
m_3^2 & = &  \frac{1}{2} \left[ \left(\delta_{31} + A \right)
           + \sqrt{\left(\delta_{31} \cos 2 \phi - A \right)^2 +
                   \left(\delta_{31} \sin 2 \phi \right)^2} \right] 
\label{eq:m3sq} 
\end{eqnarray}
Equations (\ref{eq:t2phim}), (\ref{eq:m2sq}) and (\ref{eq:m3sq})
 are valid for neutrinos. For anti-neutrinos,
we get a similar set of formulae with $A$ replaced by $-A$.

The neutrinos produced in the atmosphere enter the earth after
travelling through the atmosphere for about $20$ Km and finally
reach the detector after travelling through the earth. The distance
travelled through the earth is a function of the zenith angle.
For the five bins considered by Kamiokande \cite{fukuda}, 
the average values of
the cosine of the zenith angle are $-0.8$, $-0.4$, $0.0$, $0.4$, $0.8$ 
and the average distances travelled through the earth are $10210$,
$5137$, $832$, $34$, $6$ Km respectively \cite{giunti}.

A neutrino of flavor $\alpha$, produced in the atmosphere
at time $t = 0$, propagates
through the atmosphere as a linear combination of the vacuum mass
eigenstates. If the neutrino enters earth at time $t = t_1$, its
state vector at that time can be written as
\begin{equation}
| \Psi_{\alpha}  (t_1) \rangle = \sum_i U^v_{\alpha i} exp \left( 
-i \frac{\mu_i^2 t_1}{2 E} \right) | \nu_i^v \rangle.
\end{equation}
Reexpressing the vacuum mass eigenstates in terms of flavor
states, we have 
\begin{equation}
| \Psi_{\alpha}  (t_1) \rangle = \sum_i U^v_{\alpha i} exp \left( 
-i \frac{\mu_i^2 t_1}{2 E} \right) \sum_\lambda 
U^{v*}_{\lambda i} | \nu_\lambda \rangle.
\end{equation}
After entering the earth, the neutrino propagates as a linear 
combination of the matter dependent mass eigenstates. 
We take the earth to be a slab of constant density. 
At the time of detection $t = t_d$, 
the state vector takes the form
\begin{equation}
| \Psi_{\alpha} (t_d) \rangle = \sum_i U^v_{\alpha i} exp \left( 
-i \frac{\mu_i^2t_1}{2 E} \right) \sum_\lambda 
U^{v*}_{\lambda i} \sum_j U^m_{\lambda j} 
exp \left(-i \frac{m_j^2(t_d-t_1)}{2 E} \right) | \nu_j^m \rangle.
\end{equation}
Hence the amplitude for the neutrino 
produced as flavor $\alpha$ at t = 0 
to be detected as a neutrino of flavor 
$\beta$ at time $t_d$ is given by 
\begin{equation}
\langle \nu_\beta | \Psi_{\alpha} (t_d) \rangle = 
\sum_i \sum_\lambda \sum_j U^v_{\alpha i} U^{v*}_{\lambda i}
U^m_{\lambda j} U^{m*}_{\beta j}  
exp \left( - i\frac{\mu_i^2t_1}{2 E} \right)  
exp \left(-i \frac{m_j^2(t_d-t_1)}{2 E} \right). 
\label{eq:oscamp}
\end{equation}
The probability of oscillation $P_{\alpha \beta}$
is given by the modulus square of the above amplitude. If $t_d - t_1$  
is set equal to zero (that is if the total time of travel is equal 
to the time of travel through the atmosphere) then the expression in 
eq.(\ref{eq:oscamp}) reduces to the simple vacuum oscillation 
amplitude. The same is true if the matter effects are ignored; i.e. 
if $U^m = U^v$ and $m_i = \mu_i$.

For bins $1,2$ and $3$, the distance travelled in earth is much 
larger than the distance travelled in the atmosphere. Therefore 
$t_1$ is much smaller than $t_d$ for these bins and can be 
neglected. Neglecting $t_1$, simplifies the expressions for
oscillation probability and we obtain the expressions derived
earlier in Ref. \cite{jim}. However, for bins $4$ and $5$, the 
distance of travel in atmosphere is comparable to that in earth.
Therefore $t_1$ is of the same of order of magnitude as $t_d$
and can not be neglected. Keeping $t_1 \neq 0$ in eq.(\ref{eq:oscamp})
properly takes into account the non-adiabaticity in the abrupt
change in density when the neutrino enters earth.

\section{Calculation and Results}

\subsection{Sub-GeV Data}

First we briefly describe 
our analysis of the sub-GeV data and highlight 
the contrast between it and the analysis of the multi-GeV data. 
Matter effects are unimportant for the sub-GeV data.
If the earth is taken to be a slab of
density $5.5 \ {\rm gm/cm}^3$, the matter term $A$
for the sub-GeV neutrinos is less than  $3. 8 \times 10^{-4} 
\ eV^2$. As we will shortly see, the sub-GeV data
sets a lower limit on $\delta_{31} > 10^{-3} \ eV^2$. Hence the matter
effects can be neglected and the expressions for $P_{\alpha \beta}$  
in the sub-GeV analysis are simply the vacuum oscillation
probabilities 
\begin{eqnarray}
P^0_{\alpha \beta} & = & 
\left( U^v_{\alpha 1} U^v_{\beta 1} \right)^2 + 
\left( U^v_{\alpha 2} U^v_{\beta 2} \right)^2 + 
\left( U^v_{\alpha 3} U^v_{\beta 3} \right)^2 +  
2 \ U^v_{\alpha 1} U^v_{\alpha 2} U^v_{\beta 1} U^v_{\beta 2}
    \cos \left( 2.53 \frac{d \ \delta_{21}}{E} \right) + \nonumber \\
& & 2 \ U^v_{\alpha 1} U^v_{\alpha 3} U^v_{\beta 1} U^v_{\beta 3}
    \cos \left( 2.53 \frac{d \ \delta_{31}}{E} \right) + 
2 \ U^v_{\alpha 2} U^v_{\alpha 3} U^v_{\beta 2} U^v_{\beta 3}
    \cos \left( 2.53 \frac{d \ \delta_{32}}{E} \right),     
\label{eq:pijvac} 
\end{eqnarray}
where $d$ is the distance of travel in meters, $\delta$'s are the
mass differences in $eV^2$ and $E$ is the neutrino energy in MeV.
Because we have neglected the CP violating phase, the oscillation 
probability for the anti-neutrinos is the same as that for the 
neutrinos.
Since $\delta_{21}$ is very small, the cosine term containing it in 
eq.(\ref{eq:pijvac}) can be set equal to $1$. The other two cosine  
terms are dependent on $\delta_{31}$ (and $\delta_{32} \simeq 
\delta_{31}$), the neutrino 
energy and the distance of travel which is related to the zenith angle.  
As mentioned earlier, the double 
ratio $R$ defined in eq.(\ref{eq:defR})
does not have any zenith angle dependence for the sub-GeV data.
One can account for this if it is possible to replace the distance   
dependent terms in eq.(\ref{eq:pijvac}) by their average values.
This replacement is possible only if the average distance travelled
contains many oscillation lengths. The above condition sets a lower 
limit on the mass difference $\delta_{31} > 10^{-3} \ eV^2$  
\cite{narayan}. Note that this
lower limit is consistent with the approximation $\delta_{31} \gg
\delta_{21} \sim 10^{-5} \ eV^2$, which was made so that both solar
and atmospheric neutrino problems could be solved simultaneously.

$P^0_{\alpha \beta}$ become independent of $\delta_{31}$ and 
the neutrino energy when the distance dependent terms are replaced
by their average values. They are simply functions of the mixing
angles $\phi$ and $\psi$ alone. In this approximation, 
the expression for the double
ratio $R$ can be written as \cite{narayan}
\begin{equation}  
R = \frac{P^0_{\mu \mu} + \frac{1}{r_{MC}} P^0_{e \mu}}{
P^0_{ee} + r_{MC} P^0_{\mu e}},
\label{eq:defRsG}
\end{equation}
where $r_{MC}$ is the Monte Carlo expectation of the ratio of number 
$\mu$-like events to the number of $e$-like events. From the sub-GeV  
data of Kamiokande, we find $r_{MC} = 1.912$ and 
$R = 0.60^{+0.07}_{-0.06}\pm 0.05$ \cite{hirata,narayan}.
We take the allowed values of 
$\phi$ from our analysis of solar neutrino data which 
restricts $\phi$ to be in the range 
$0 \leq \phi \leq 50^0$ \cite{narayan}. 
We find the allowed region in the $\phi-\psi$ plane by requiring 
the theoretical value calculated from 
eq.(\ref{eq:defRsG}) to be within 
$1 \sigma$ or $1.6 \sigma$ uncertainty of the experimental value. 
In our previous 
analysis we restricted the range of 
$\psi$ to be $0 \leq \psi \leq 45^0$. 
However, given the assumptions 
we made about vacuum mass eigenvalues, the
most general possibility is to 
allow $\psi$ to vary between $0$ and $90^0$ \cite{thankjim}.
The new region allowed by the sub-GeV data, where 
$\phi$ was restricted by the solar 
neutrino data and $\psi$ is allowed to
vary between $0$ to $90^0$, is shown in Figure 1. The region between 
the solid lines is the parameter space  
which satisfies the experimental constraints at 
$1 \sigma$ level whereas the region between dashed lines satisfies 
the experimental constraints at $1.6 \sigma$ level. 
One important point to be noted in this analysis is that the sub-GeV
data place only lower bound on $\delta_{31}$. 
The allowed region in $\phi-\psi$ plane is quite large.
%In particular, the value $\phi = 0$
%is allowed at $1 \sigma$ level. This situation is to be contrasted 
%with the bounds we get from multi-GeV data.

\subsection{Multi-GeV data}

The multi-GeV data of Kamiokande have been 
presented for five zenith angle   
bins in Ref. \cite{fukuda}. 
For each of these bins, the observed numbers of  
electron-like events and muon-like events and their Monte Carlo 
expectations (without neutrino oscillations) have been  
given. From these one can calculate two sets of ratios
\begin{eqnarray}
r^i_{MC} & = & \left( \frac{N^i_\mu}{N^i_e} \right)_{MC}  
\label{eq:defriMC}\\
r^i_{obs} & = & \left( \frac{N^i_\mu}{N^i_e} \right)_{obs}, 
\label{eq:defriobs}
\end{eqnarray}
and the set of double ratios
\begin{equation}
R^i_{obs} = r^i_{obs}/r^i_{MC} \label{eq:defRi}
\end{equation}
for each bin $i = 1,2,...,5$.
We summarize the multi-GeV data of Kamiokande \cite{fukuda} in Table I.
 
The $\mu$-like events are subdivided into fully contained (FC) and
partially contained (PC) events whereas all the $e$-like events are
fully contained. The efficiency of detection for each type of event 
is different and is a funtion of the neutrino energy. Thus we have 
three detection efficiencies $\varepsilon^{\mu}_{FC} (E)$, 
$\varepsilon^{\mu}_{PC} (E)$ and $\varepsilon^e (E)$. We obtained 
these efficiencies from Kamiokande Collaboration \cite{kajita}. 
  
The expected number of $\mu$-like and $e$-like events, in the absence  
of neutrino oscillations, is given by 
\begin{eqnarray}
N^i_{\mu}|_{M.C.} & = & 
\int \left[ \phi^i_{\nu_{\mu}} (E) \sigma (E)
+ \phi^i_{\bar{\nu}_{\mu}} (E) \bar{\sigma} (E) \right] 
\left( \varepsilon^{\mu}_{FC} (E) + 
\varepsilon^{\mu}_{PC} (E) \right) dE 
\label{eq:nimuno}  \\
N^i_e|_{M.C.} & = & 
 \int \left[ \phi^i_{\nu_e} (E)  \sigma (E) +  
\phi^i_{\bar{\nu}_e} (E) \bar{\sigma} (E) \right] 
\varepsilon^e (E) dE  
\label{eq:nieno}  
\end{eqnarray}
where $\phi$'s are the fluxes of the atmospheric neutrinos at the
location of Kamiokande. These are tabulated in Ref. \cite{honda} as
functions of the neutrino energy (from $E = 1.6$ GeV to $E = 100$
GeV) and the zenith angle. $\sigma$ and $\bar{\sigma}$
are the charged current cross sections of neutrinos and  
anti-neutrinos respectively with nucleons.  The cross section is
the sum of the quasi-elastic scattering and the deep inelastic
scattering (DIS). The values for quasi-elastic scattering are 
taken from
Gaisser and O'Connel \cite{gaisser} and those for the DIS are taken
from Gargamelle data \cite{musset}. In calculating the DIS cross
section, we took the lepton energy distribution to be given by the
scaling formula (which is different for neutrinos and anti-neutrinos)
and integrated $d \sigma/d E_{lep}$ 
from the minimum lepton energy $E_{min}
= 1.33$ GeV to the maximum lepton energy $E_{max} = E_\nu - m_\pi$.  
The maximum lepton energy is chosen by defining DIS to contain at  
least one pion in addition to the charged lepton and the baryon.
The differences in the fiducial volumes and exposure times for 
fully contained and partially contained events have been incorporated
into the detection efficiency $\varepsilon^\mu_{PC} (E)$. 
From equations (\ref{eq:nimuno}) and (\ref{eq:nieno}) we calculate
our estimation of the Monte Carlo expectation of $r^i_{MC}$, 
the ratio of the $\mu$-like events to the $e$-like events. 
The numbers we obtain are within $10 \%$ of the values quoted by  
Kamiokande collaboration in Ref. \cite{fukuda}. The differences
could be due to the different set fluxes used \cite{fn2} and due
to the simple approximation we made for the cross sections. 

In the presence of oscillations, the number of $\mu$-like 
and $e$-like events are given by 
\begin{eqnarray}
N^i_{\mu}|_{osc} & = & 
\int \left[ \phi^i_{\nu_{\mu}} P_{\mu \mu} \sigma
+ \phi^i_{\bar{\nu}_{\mu}} P_{\bar{\mu} \bar{\mu}} 
   \bar{\sigma}  
+ \phi^i_{\nu_e} P_{e \mu} \sigma
+ \phi^i_{\bar{\nu}_e} P_{\bar{e} \bar{\mu}} 
   \bar{\sigma} \right] 
\left( \varepsilon^\mu_{FC} + \varepsilon^\mu_{PC} \right) dE  
\label{eq:nimuos} \\
N^i_e|_{osc} & = & 
\int \left[ \phi^i_{\nu_e} P_{ee} \sigma
+ \phi^i_{\bar{\nu}_e} P_{\bar{e} \bar{e}} 
   \bar{\sigma} 
+ \phi^i_{\nu_{\mu}} P_{\mu e} \sigma
+ \phi^i_{\bar{\nu}_{\mu}} P_{\bar{\mu} \bar{e}} 
   \bar{\sigma} \right] \varepsilon^e dE \label{eq:nieos}
\end{eqnarray}
where $P_{\alpha \beta}$s are the probabilities for neutrino of flavor 
$\alpha$ to oscillate into flavor $\beta$, derived in the last section.
These oscillation probabilities are functions of the distance of travel
$d$, the mixing angles $\phi$ and $\psi$, the mass difference 
$\delta_{31}$ and the matter term $A$. These are calculated using
the formulae in eqs. (\ref{eq:t2phim}), 
(\ref{eq:m2sq}), (\ref{eq:m3sq}) and (\ref{eq:oscamp}).
The probability of oscillation for anti-neutrinos 
$P_{\bar{\alpha} \bar{\beta}}$, in general, is different from
$P_{\alpha \beta}$ because of the different $A$ dependence. 

From equations (\ref{eq:nimuos}) and (\ref{eq:nieos}) we can 
calculate the ratio of $\mu$-like events to $e$-like events in the
presence of oscillations to be
\begin{equation}
r^i_{osc} = \left( \frac{ N_\mu^i}{N_e^i} \right)_{osc}
\label{eq:defriosc}
\end{equation}
and the double ratio
\begin{equation}
R^i_{osc} = \frac{r^i_{osc}}{r^i_{MC}}. \label{eq:defRiosc}
\end{equation} 

If the atmospheric neutrino deficit is due to neutrino oscillations,
then the double ratios $R^i_{osc}$ given in eq. (\ref{eq:defRiosc}) 
should be within the range of the corresponding observed double ratios
$R^i_{obs}$, which are given in Table I. We searched for the values 
of the neutrino parameters $\phi, \psi$ and $\delta_{31}$
for which the predicted values of $R^i_{osc}$  
satisfy the experimental constraints on the double ratios for all the
five bins. The ranges of variation in the three parameters are
\begin{enumerate}
\item
$0 \leq \phi \leq 50^0$. This is the range of $\phi$ allowed by the
solar neutrino problem. For this range of $\phi$, there exist values
of $\delta_{21}$ and $\omega$ such that all the three solar neutrino
experiments can be explained \cite{narayan,fogli3}.
\item
$0 \leq \psi \leq 90^0$. $\psi$ is varied over its fully allowed range.
\item
$10^{-3} \ eV^2 \leq \delta_{31} \leq 10^{-1} \ eV^2$. The lower limit
is given by the sub-GeV data and the upper limit is the largest value
allowed by the two flavor analysis of the multi-GeV data by    
Kamiokande \cite{fukuda}.
\end{enumerate}
The results are plotted in Figures 2, 3 and 4. 
Figure 2 gives the projection of the 
allowed region on the $\phi-\psi$ plane, Figure 3 gives the projection
on the $\phi-\delta_{31}$ plane and Figure 4 gives the projection on
the $\psi-\delta_{31}$ plane. 
The solid lines enclose
the regions of parameter space whose predictions lie within the 
experimental range given by $1 \sigma$ uncertainties. The broken lines
enclose regions whose predictions fall within range given by 
$1.6 \sigma$ uncertainties.

As seen from Table I, 
the uncertainties in bin $5$, which has $\langle \cos \theta \rangle
= 0.8$, are quite large compared to the uncertainties in the other
four bins. Morevoer, $r^5_{obs}$ is greater than $r^5_{MC}$ through 
most of its range. Hence the double ratio $R^5_{obs} > 1$ for most of
its range. The Monte Carlo expectation of the electron neutrino flux
is less than that of the muon neutrino flux and the oscillation 
probabilities $P_{\alpha \beta}$ are all less than $1$. Using these
facts, one can show from eqs. 
(\ref{eq:defriMC}) and (\ref{eq:defriosc})
that, in general, $r^i_{osc} \leq r^i_{MC}$ or $R^i_{osc} \leq 1$.
Therefore, the region of overlap between $R^5_{obs}$ and $R^5_{osc}$   
is very small. It is $0.9 - 1.0$ for 
$1 \sigma$ uncertainties. It is possible that this small overlap 
is imposing a very strong constraint, 
leading to a situation where the bin with the
largest uncertainty is essentially controlling the allowed values 
of the parameters. Because of this unsatisfactory situation, we redid
the analysis ignoring the constraint from bin $5$. We searched for 
regions of parameter space for which the values of $R^i_{osc}$ are
within the ranges of corresponding $R^i_{obs}$ for only the first four
bins.  

The results of the $4$ bin analysis are  plotted in Figures 5, 6 and 7. 
Figure 5 gives the projection of the 
allowed region on the $\phi-\psi$ plane, Figure 6 gives the projection
on the $\phi-\delta_{31}$ plane and Figure 7 gives the projection on
the $\psi-\delta_{31}$ plane. 
As before, the solid lines enclose the regions satisfying $1 \sigma$
vetoes and the broken lines enclose regions allowed by $1.6 \sigma$ 
vetoes. Comparing the corresponding figures we find that the allowed
values of parameters for the $4$ bin fit are the almost identical to
those from the $5$ bin fit at $1.6 \sigma$ level. 
The allowed regions at $1 \sigma$ level are somewhat larger
compared to the $5$ bin fit. This is not surprising because 
the bin $5$ imposes a very strong constraint at $1 \sigma$ level.
If this constraint is relaxed, then a somewhat larger region is
allowed. The $4$ bin fit shows that the $5$th bin, which
has the largest uncertainty, does not exercise undue 
influence on the selection of the parameter space.
 
\section{Discussion And Conclusions}

The parameter space shown in Figures 1 to 4 , together with the 
allowed values for $\omega$ and $\delta_{21}$ from our earlier work
\cite{narayan}, provides a complete solution to the solar and the
atmospheric neutrino problems in the framework of three flavor 
neutrino oscillations. 
The salient features of the results of the multi-GeV analysis are:
\begin{itemize}
\item
Most of the parameter space allowed by the multi-GeV data is a subset 
of the space allowed by the sub-GeV data.
\item
The range of $\delta_{31}$ allowed by $1 \sigma$ vetoes is extremely
narrow. It is very close to the best fit value 
given by the two flavor analysis of Kamiokande.
\item
The value of $\psi$ is always large $(\psi \geq 40^0)$
and $\psi = 90^0$ is allowed.
\item
In the region allowed by $1 \sigma$ vetoes $\phi$ is always
non-zero. $\phi = 0$ is allowed only at $1.6 \sigma$ vetoes.
\end{itemize}

From the parametrization of $U^v$, effective two level mixings can be 
obtained for the following choices of the angles:
\begin{itemize}
\item 
$\nu_e \leftrightarrow \nu_\mu$ for $\psi = 90^0$,
\item 
$\nu_e \leftrightarrow \nu_\tau$ for $\psi = 0$ and
\item 
$\nu_\mu \leftrightarrow \nu_\tau$ for $\phi = 0$. 
\end{itemize}
Any solution to the atmospheric neutrino problem should suppress 
the muon neutrinos, enhance the electron neutrinos or do both. 
The $\nu_e \leftrightarrow \nu_\tau$ channel, which suppresses 
electron neutrinos but leaves muon neutrinos untouched, cannot
account for the atmospheric neutrino problem. Hence any solution 
of atmospheric neutrino problem should be away from the effective
two flavor $\nu_e \leftrightarrow \nu_\tau$ oscillations. The 
large value of the angle $\psi$ is just a reflection of this fact.
The allowed region includes the value $\psi = 90^0$. 
Then the atmospheric neutrino problem is explained purely in terms of 
the two flavor oscillations between $\nu_e \leftrightarrow \nu_\mu$,
with the relevant mass difference being $\delta_{31}$. In this case,
the solar neutrino problem is solved by $\nu_e \rightarrow \nu_\tau$ 
oscillation which is determined by the mass difference $\delta_{21}$
(and the mixing angles $\omega$ and $\phi$).

%From figures 2(d),(e) and (f) of ref. \cite{fukuda} it can be seen
%that most of the expected multi-GeV events are caused by neutrinos
%with energies less than 10 GeV (over $80 \%$ for muon-like events
%and over $90 \%$ for electron-like events). The matter term, for an
%energy of 5 GeV, is about $2 \times 10^{-3} eV^2$. This is much 
%smaller than the $\delta_{31}$ that we obtain from our analysis.
%Therefore, it is likely that matter effects may not play an 
%important role in determining the allowed parameter regions in 
%the analysis of the multi-GeV data. To check this, we reran our
%programs with the matter term set equal to zero. With this change,
%the double ratio $R^i_{osc}$, defined in equation           
%(\ref{eq:defriosc}), changes by about $10 \%$ in the first bin
%and by about $5 \%$ in the second bin. There is no discernible
%change in the other three bins. Since the errors in $R^i_{obs}$
%are about $30 \%$, these small changes in $R^i_{osc}$ do not lead 
%to any change in the allowed regions of the parameter space. 

How important are the matter effects in the analysis of the
multi-GeV data? It can be seen from figures 2(d)-2(f) of Ref.
\cite{fukuda} that most of the expected multi-GeV events are caused 
by neutrinos with energies less than 10 GeV (over $80 \%$ for 
muon-like events and over $90 \%$ for electron-like events). 
The matter term, for a neutrino of energy 5 GeV, is about
$2 \times 10^{-3} eV^2$. Since the initial range we considered for
$\delta_{31}$ varied from $10^{-3} \ eV^2$ to $0.1 \ eV^2$, apriori
one must include the matter effects in the expressions for the 
oscillation probabilities. However, the value of $\delta_{31}$ 
in the allowed region, especially for the $1 \sigma$ vetoes,
where it is about $0.03 \ eV^2$, is much 
larger than the matter term. Therefore, it is likely that the 
matter effects may not play an important role in determining the 
allowed parameter regions in the analysis of the multi-GeV data.
To check this we reran our program with the matter term set equal
to zero. With this change, the double ratios $R^i_{osc}$, defined 
in equation (\ref{eq:defriosc}), changes by about $10 \%$ in the 
first bin and by about $5 \%$ in the second bin. There is no 
discernible change in the other three bins. Since the errors in 
$R^i_{obs}$ are about $30 \%$, these small changes in $R^i_{osc}$  
do not lead to any apprecialble change in the allowed regions of the  
parameter space. However, the effect of matter terms may become  
discerible when more accurate data from Super Kamiokande become 
available.

Since the earth matter effects seem to play no role in the
determination of the parameter space, can one interpret the 
observed zenith angle dependence purely in terms of vacuum 
oscillations? For an energy of 5 GeV, the mass square difference
$10^{-2} \ eV^2$ corresponds to an oscillation length of about 
$1200$ Km. Thus bins $1$ and $2$ contain many oscillation lengths
and the second and the third cosine terms in the vacuum oscillation 
probability, given in equation (\ref{eq:pijvac}), average out to
zero. For bins $4$ and $5$ the cosine terms are almost 1. In bin 
$3$, these terms take some intermediate value. Therefore we have
large suppression in the first two bins, almost no suppression
in the last two bins and moderate suppression in the middle bin.

Eventhough matter terms play no important role in the multi-GeV
analysis, $\nu_e \leftrightarrow \nu_\mu$ oscillations provide
a better fit to data compared to $\nu_\mu \leftrightarrow 
\nu_\tau$ oscillations. In $\nu_\mu \leftrightarrow \nu_\tau$
oscillations, the $\nu_\mu$ flux is suppressed whereas the 
$\nu_e$ flux is untouched. Thus the double ratios $R^i_{osc}$
do become less than $1$ but the large suppression observed in
bins $1$ and $2$ is difficult to achieve via this channel. 
For $\nu_e \leftrightarrow \nu_\mu$ oscillations, the $\nu_\mu$
flux is reduced and $\nu_e$ flux is increased. This occurs even
when $P_{\mu e} = P_{e \mu}$, which is the case for vacuum 
oscillations, because the flux of $\nu_\mu$ is roughly twice the
flux of $\nu_e$ before oscillations. Hence the double ratios 
$R^i_{osc}$ are smaller for the case of $\nu_e \leftrightarrow
\nu_\mu$ oscillations compared to the case of $\nu_\mu
\leftrightarrow \nu_\tau$ oscillations and they fit the data 
better.

We now make a brief comment on the LSND results on the search for
$\bar{\nu}_{\mu} \rightarrow \bar{\nu}_e$ oscillations
\cite{lsnd} in the context of our 
analysis of atmospheric neutrinos.
The LSND collaboration gives an oscillation probability 
$P_{\bar{\mu} \bar{e}} 
= (3.1^{+1.1}_{-1.0} \pm 0.5) \times 10^{-3}$ for muon anti-neutrinos
in the energy range $20 - 60$ MeV. In the framework described in 
section II, the oscillation probability relevant for the LSND
experiment is the vacuum oscillation probability 
\begin{equation}
P^0_{\bar{\mu} \bar{e}} = P^0_{\mu e} =
\sin^2 2 \phi \sin^2 \psi \sin^2 \left(1.27 \frac{d \ \delta_{31}}{E}
\right) \label{eq:lsndosc}.
\end{equation} 
Note that both $\phi$ and $\psi$ have to be non-zero for $P_{\mu e}$ to
be non-zero. In the region allowed by the $1 \sigma$ vetoes of the 
multi-GeV atmospheric neutrino data we have 
\begin{eqnarray}
{\rm Minimum} \ \left( \sin^2 2 \phi \sin^2 \psi \right) \simeq 0.04
& \ {\rm for} \ & \phi \simeq 8^0, \psi \simeq 40^0 \nonumber \\ 
{\rm Maximum} \ \left( \sin^2 2 \phi \sin^2 \psi \right) \simeq 1
& \ {\rm for} \ & \phi \simeq 40^0, \psi \simeq 90^0. \nonumber  
\end{eqnarray}
Substituting these values and the oscillation probability obtained by
LSND in eq. (\ref{eq:lsndosc}), we obtain
\begin{equation}
0.001 \leq 
\sin^2 \left(1.27 \frac{d \ \delta_{31}}{E} \right)
\leq 0.1. 
\end{equation} 
For the LSND experiment the distance $d = 30$ meters. Taking the 
average energy to be $\langle E \rangle = 40$ MeV, we obtain the
range of $\delta_{31}$ to be 
\begin{equation}
0.03 \ eV^2 \leq \delta_{31} \leq 0.3 \ eV^2. \label{eq:lsndran}
\end{equation} 
From the analysis of multi-GeV atmospheric neutrino data we have
the upper limit on $\delta_{31} \leq 0.06 \ eV^2$ (Figures 3 and 4). 
Hence there is a small region of overlap between the range of 
neutrino parameters required by the atmospheric neutrino data and
the LSND data. This suggests that the standard three flavor analysis
can accommodate all the data so far \cite{cafu} and perhaps a fourth
sterile neutrino is not needed. 

In conclusion, we have analyzed the atmospheric neutrino data of 
Kamiokande in the context of three flavor neutrino oscillations.
We took into account both the zenith angle dependence of the 
multi-GeV data and the matter effects due to the propagation of 
the neutrinos through the earth. We obtained the regions in 
neutrino parameters which solve both the solar and the atmospheric
neutrino problems. We found that the matter effects have negligible
influence on atmopsheric neutrinos even in the multi-GeV range. The
allowed regions of the parameter space with or without matter 
effects are almost identical. 

\bigskip

As we finished this work, we came across a preprint by Fogli 
{\it et al} \cite{fogli4}. They have analyzed the sub-GeV data
from various experiments and the zenith angle dependent multi-GeV
data from Kamiokande. Although a detailed comparison is difficult, 
qualitatively our results agree with theirs.
\bigskip

Acknowledgements: We are grateful to M. V. N. Murthy for collaboration 
during the earlier part of this work and for a critical reading
of the manuscript. We are specially grateful to Prof. Kajita of 
the Kamiokande Collaboration for supplying us with the detection
efficiencies of the Kamiokande detector. We thank S. R. Dugad, 
Rahul Sinha, N. K. Mondal and K. V. L. Sarma for numerous discussions
during the course of this work. We also thank
Jim Pantaleone, Sandip Pakvasa and the referee for critical comments.

\begin{table}
\begin{tabular}{|c|c|c|c|c|c|}
\hline
Bin No. & $\langle cos \theta \rangle$ 
& $\langle distance \rangle$ in Km
& $r^i_{MC}$ & $r^i_{obs}$ & $R^i_{obs}$ \\ \hline
1 & -0.8 & 10,210 & 3.0 & $0.87^{+0.36}_{-0.21}$ &  
$0.29^{+0.12}_{-0.07}$ \\ \hline 
2 & -0.4 &  5,137 & 2.3 & $1.06^{+0.39}_{-0.30}$ &  
$0.46^{+0.17}_{-0.13}$ \\ \hline
3 &  0.0 &    832 & 2.1 & $1.07^{+0.32}_{-0.23}$ &  
$0.51^{+0.15}_{-0.11}$ \\ \hline
4 &  0.4 &     34 & 2.3 & $1.45^{+0.51}_{-0.34}$ &  
$0.63^{+0.22}_{-0.16}$  \\ \hline
5 &  0.8 &      6 & 3.0 & $3.9^{+1.8}_{-1.2}$ & 
$1.3^{+0.6}_{-0.4}$   \\ \hline 
\end{tabular}
\caption{Zenith angle dependent data from Kamiokande [6]}
\end{table}

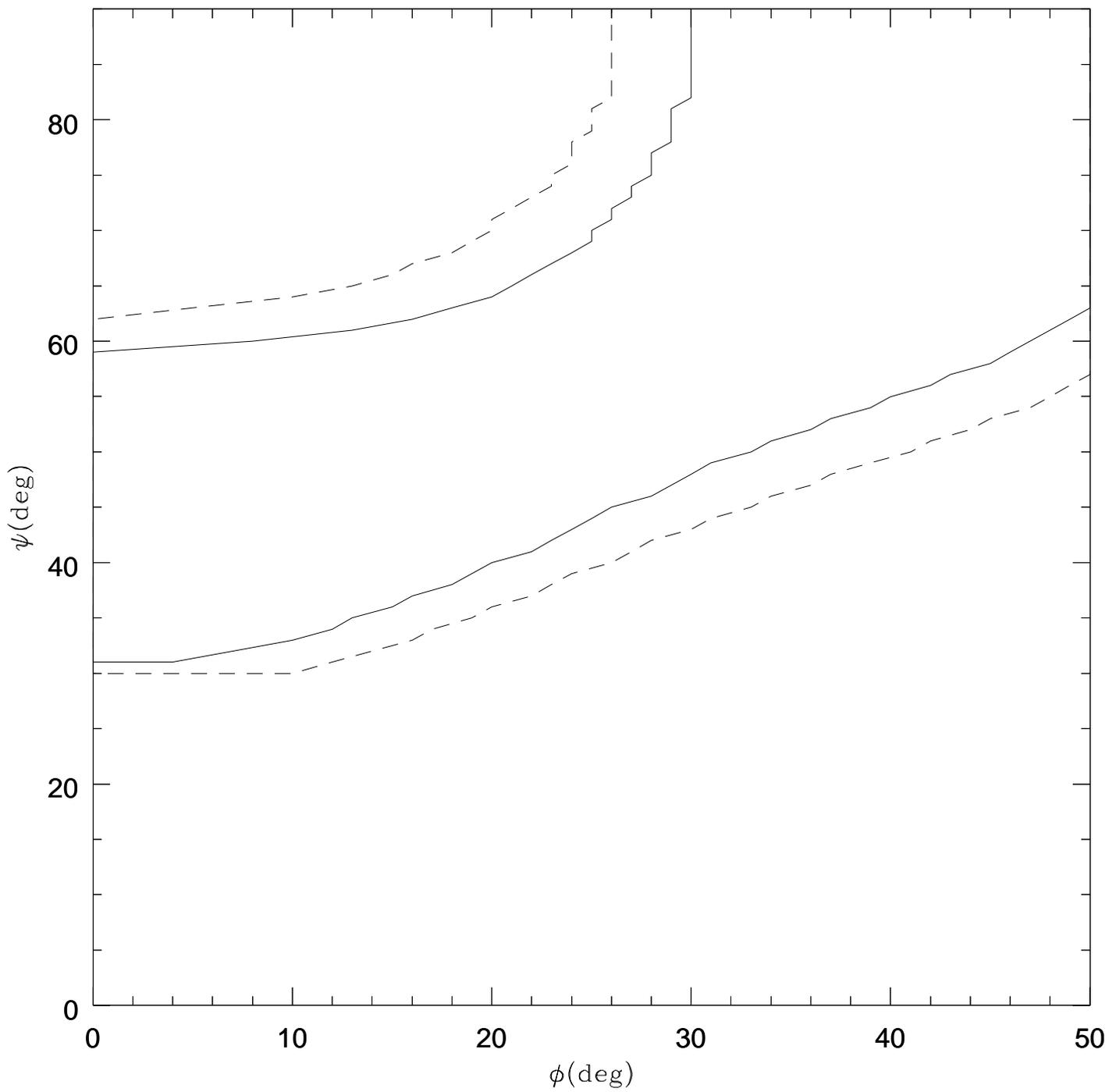
\begin{figure}
\caption{Allowed region in $\phi-\psi$ plane by the sub-GeV data
$( {\rm with} \ \delta_{31} \geq 10^{-3} \ eV^2)$ at $1 \sigma$
(enclosed by solid lines) and at $1.6 \sigma$ (enclosed by broken
lines) 
}\label{Fig. 1}
\end{figure}

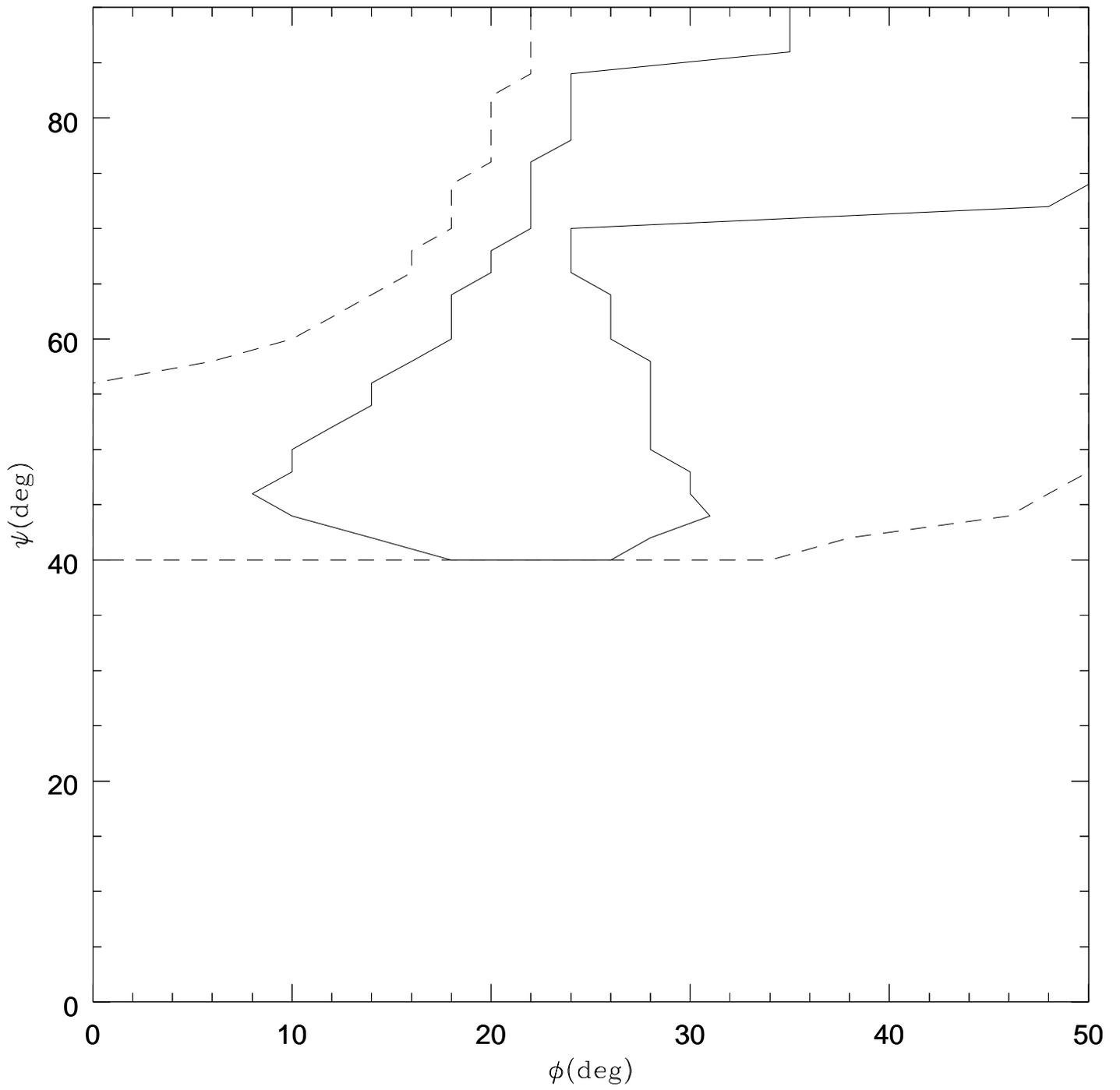
\begin{figure}
\caption{Allowed region in $\phi-\psi$ plane by $5$ bin analysis 
of multi-GeV data
$( {\rm with} \ 10^{-3} \ eV^2 \leq \delta_{31} \leq 10^{-1} \ eV^2)$ 
at $1 \sigma$ (enclosed by solid lines) and 
at $1.6 \sigma$ (enclosed by broken lines)
}\label{Fig. 2}
\end{figure}

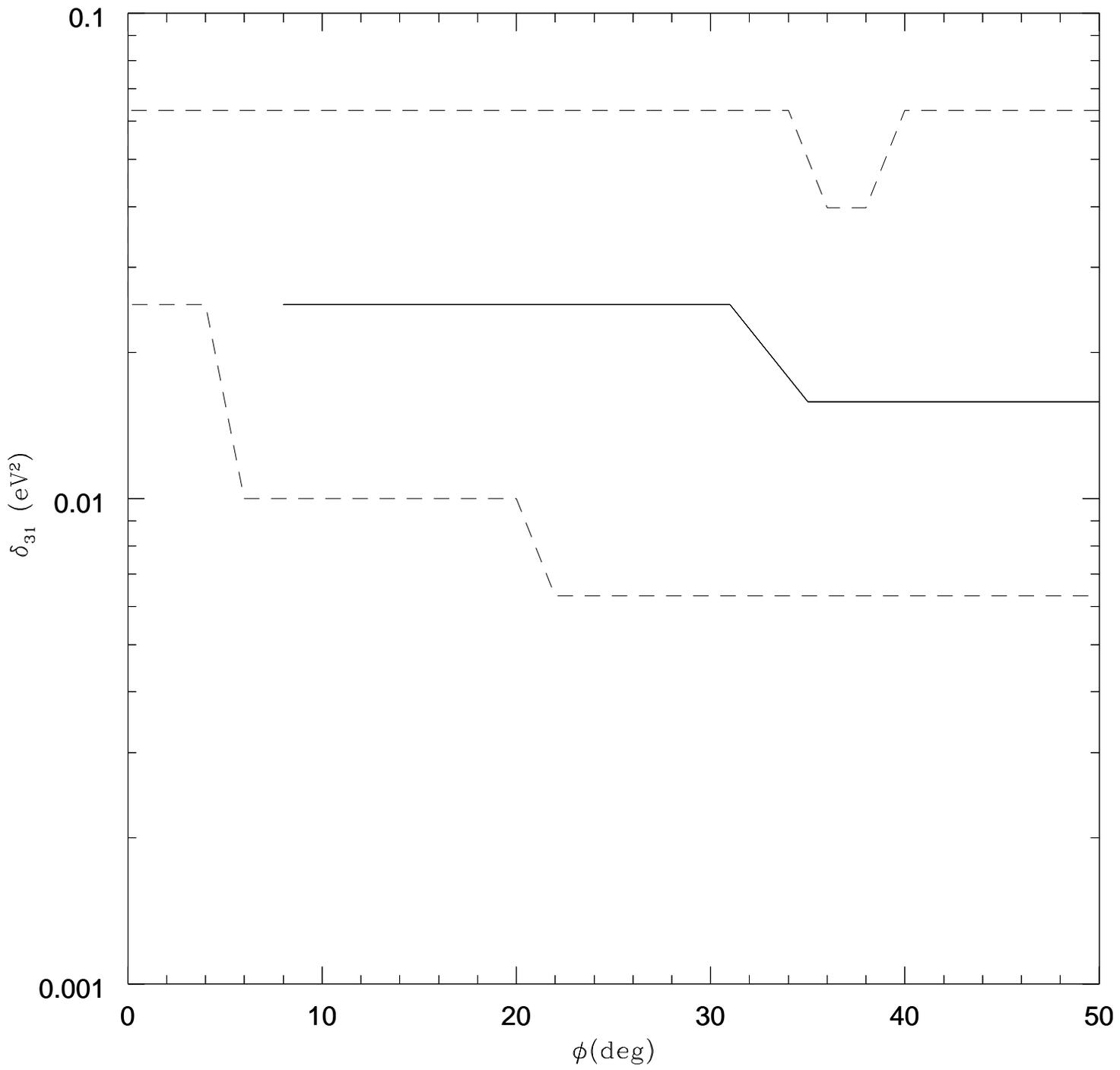
\begin{figure}
\caption{Allowed region in $\phi-\delta_{31}$ plane by $5$ bin 
analysis of multi-GeV data 
$({\rm with} \ 0 \leq \psi \leq 90^0)$ 
at $1 \sigma$ (enclosed by solid lines) and 
at $1.6 \sigma$ (enclosed by broken lines)
}\label{Fig.3}
\end{figure}

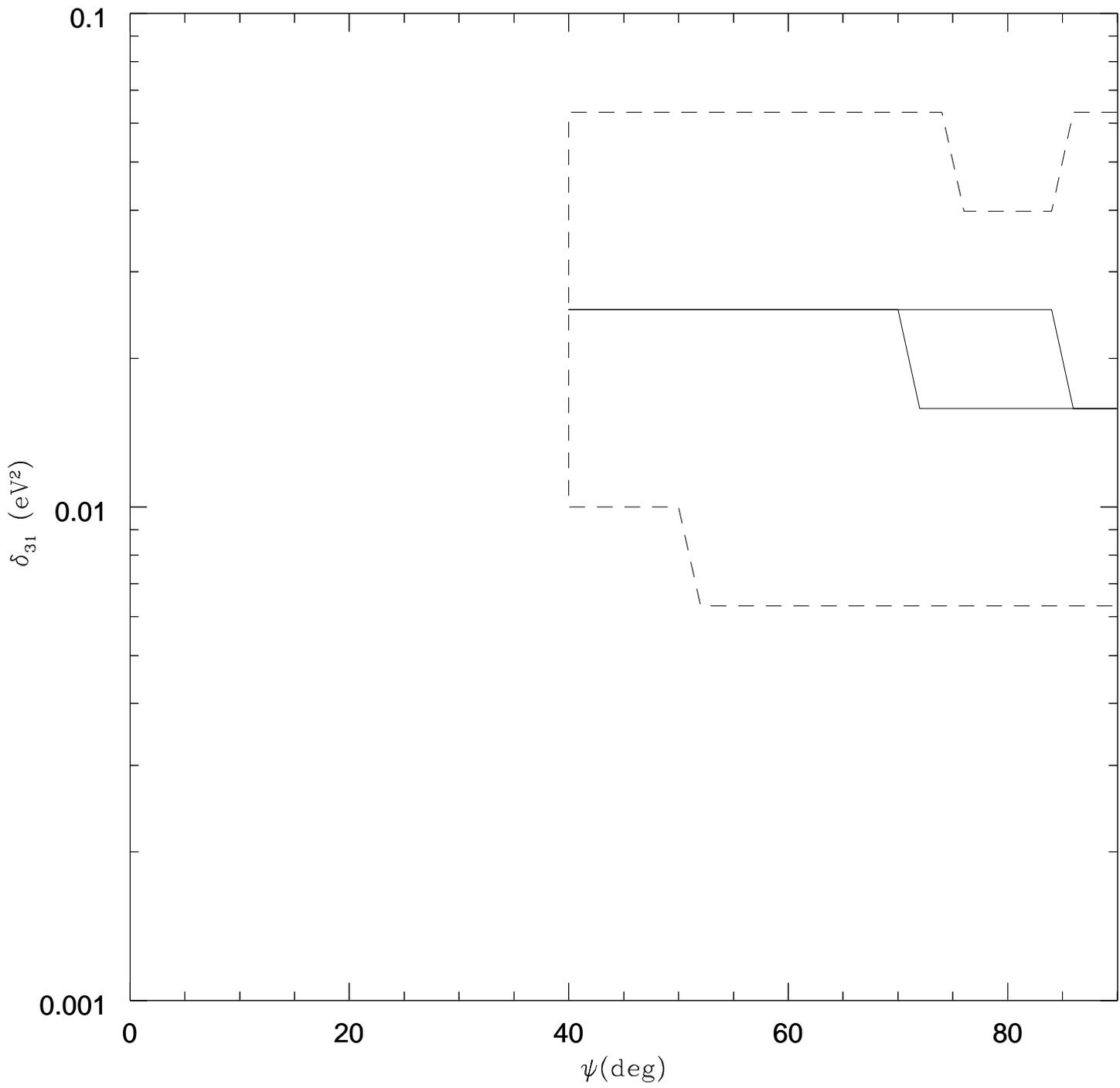
\begin{figure}
\caption{Allowed region in $\psi-\delta_{31}$ plane by $5$ bin 
analysis of multi-GeV data 
$({\rm with} \ 0 \leq \phi \leq 50^0)$ 
at $1 \sigma$ (enclosed by solid lines) and 
at $1.6 \sigma$ (enclosed by broken lines)
}\label{Fig.4}
\end{figure}

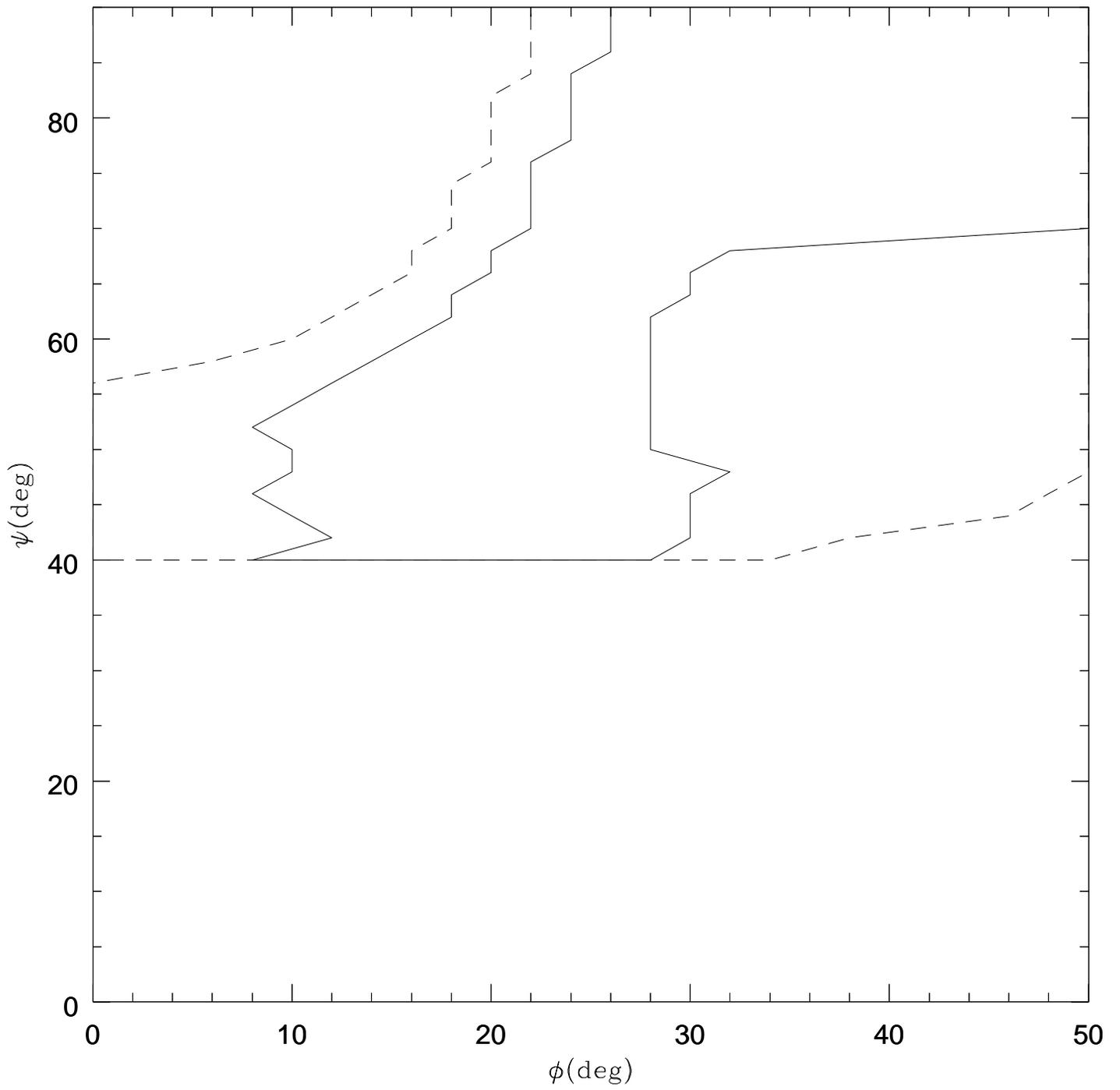
\begin{figure}
\caption{Allowed region in $\phi-\psi$ plane by $4$ bin analysis 
of multi-GeV data
$( {\rm with} \ 10^{-3} \ eV^2 \leq \delta_{31} \leq 10 \ eV^2)$ 
at $1 \sigma$ (enclosed by solid lines) and 
at $1.6 \sigma$ (enclosed by broken lines)
}\label{Fig. 5}
\end{figure}

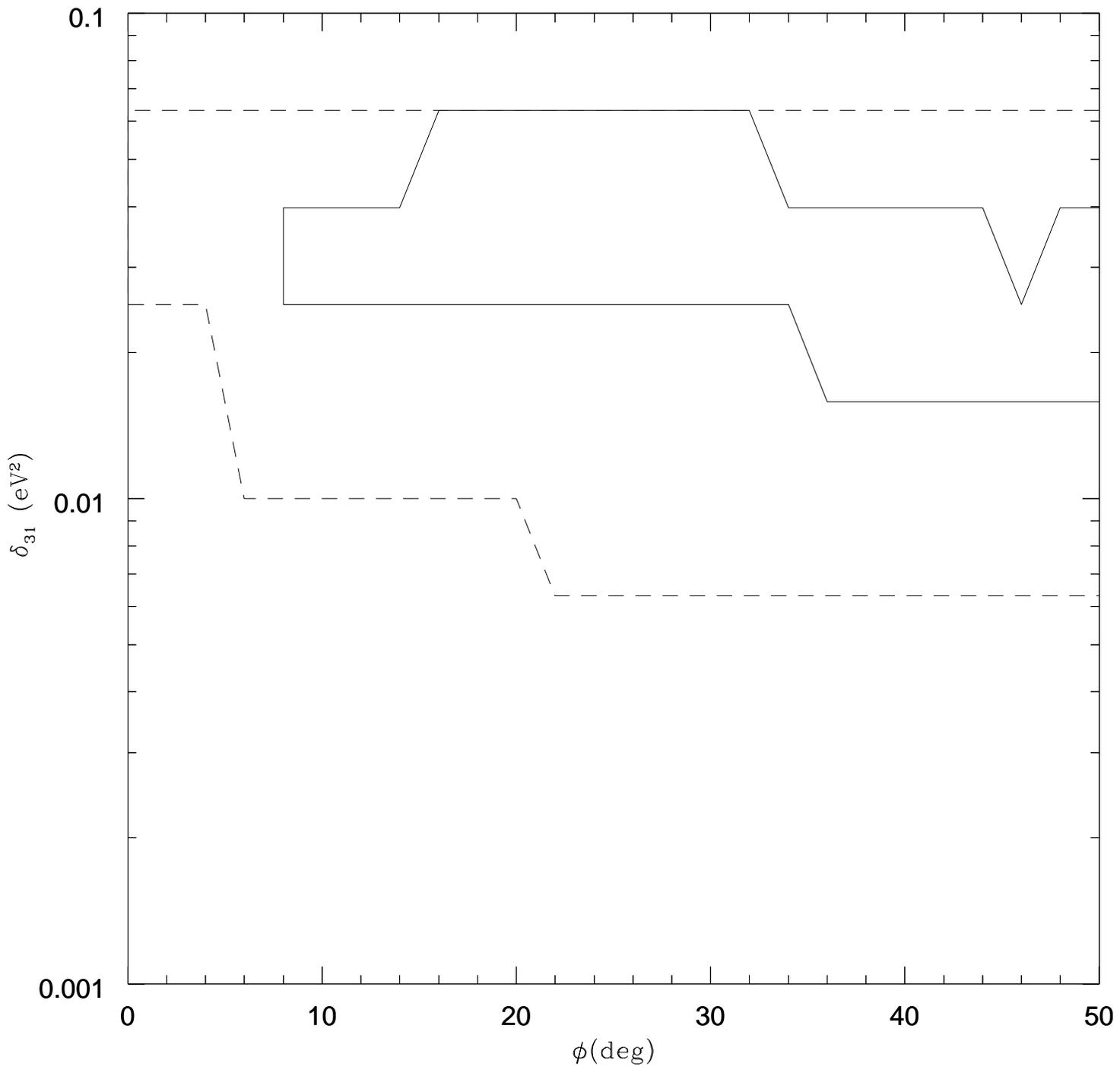
\begin{figure}
\caption{Allowed region in $\phi-\delta_{31}$ plane by $4$ bin 
analysis of multi-GeV data 
$({\rm with} \ 0 \leq \psi \leq 90^0)$ 
at $1 \sigma$ (enclosed by solid lines) and 
at $1.6 \sigma$ (enclosed by broken lines)
}\label{Fig.6}
\end{figure}

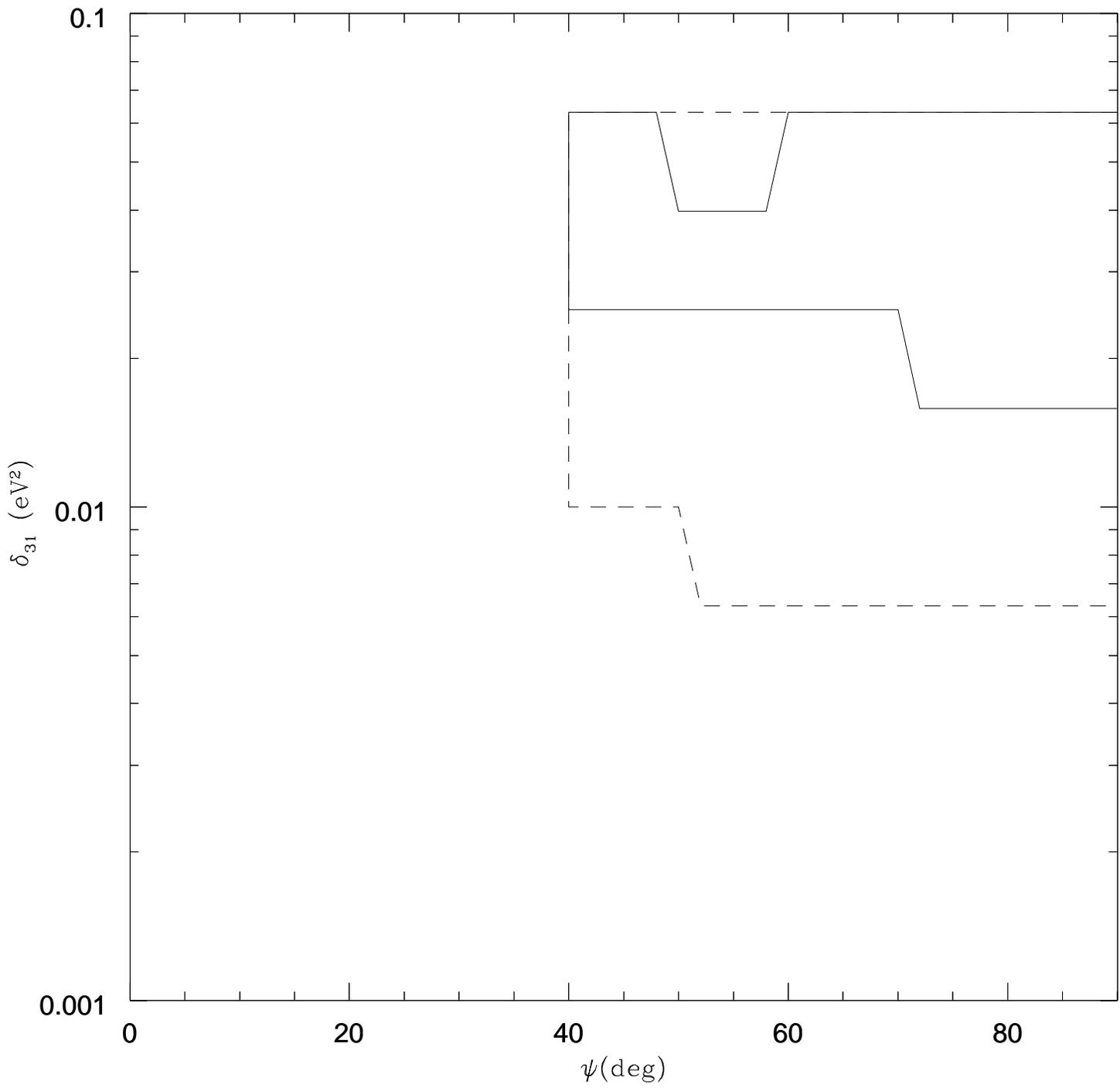
\begin{figure}
\caption{Allowed region in $\psi-\delta_{31}$ plane by $4$ bin 
analysis of multi-GeV data 
$({\rm with} \ 0 \leq \phi \leq 50^0)$ 
at $1 \sigma$ (enclosed by solid lines) and 
at $1.6 \sigma$ (enclosed by broken lines)
}\label{Fig.7}
\end{figure}

\end{document}